\documentclass[intlimits,twoside,a4paper]{article}

\usepackage[cp1251]{inputenc}
\usepackage[eqsecnum]{cmpj3}

\usepackage{bm}

\usepackage{color}

\issue{2023}{26}{2}{23703}
\doinumber{10.5488/CMP.26.23703}

\title {Dirac fermion spectrum of the fractional quantum Hall states 
}
\author[I. N. Karnaukhov]{I. N. Karnaukhov\orcid{0000-0002-1722-070X}\thanks{Corresponding author: \email{karnaui@yahoo.com}.}}
\address{G. V. Kurdyumov Institute for Metal Physics of the
	NAS of Ukraine, 36 Vernadsky Boulevard, 03142 Kyiv, Ukraine}

\Keywords{Harper--Hofstadter model, quantum Hall effect, fractional quantum Hall effect}

\date{Received October 20, 2022, in final form January 09, 2023}

\begin{document}
	
	\maketitle

\begin{abstract}
Applying a unified approach, we study the integer quantum Hall effect (IQHE) and  fractional quantum Hall effect~(FQHE) in the Hofstadter model with short range interactions between fermions. An effective field, that takes into account the interaction between fermions, is determined by both amplitude and phase. Its amplitude is proportional to the interaction strength, the phase corresponds to the minimum energy. In fact, the problem is reduced to the Harper equation with two different scales: the first is a magnetic scale with the cell size corresponding to a unit quantum magnetic flux, the second scale determines the inhomogeneity of the effective field, forms the steady fine structure of the Hofstadter spectrum and leads to the realization of  fractional quantum Hall states.
In a sample of finite size with open boundary conditions, the fine structure of the Hofstadter spectrum consists of the Dirac branches of the fermion excitations and includes the fine structure of the edge chiral modes.
The Chern numbers of the topological Hofstadter bands are conserved during the formation of their  fine structure.  The edge modes are formed into the Hofstadter bands. They connect the nearest-neighbor subbands and determine the conductance for the fractional filling.
\printkeywords
\end{abstract}
\maketitle

\section{Introduction}

The Harper--Hofstadter model~\cite{Har,H} plays a key role in the modern understanding and description of topological states on a 2D lattice. It allows us to describe the nontrivial behavior of fermions in an external magnetic field with their arbitrary dispersion at different filling, to determine the structure of topological bands, and to calculate the Chern numbers in a wide range of magnetic fluxes. For rational magnetic fluxes penetrating into a magnetic cell with size $q$ ($q$ is defined in units of the lattice spacing), the Hofstadter model has an exact solution~\cite{w,w1}. In experimental realizable magnetic fields, which corresponds to semi-classical limit with a magnetic scale $q\simeq 10^3$--$10^4$, the spectrum of quasiparticle excitations  is well described in the framework of the Landau levels near the edge spectrum~\cite{K1,K2,K3}, and the Dirac levels in graphene~\cite{a3,a4,K4}. Irrational magnetic fluxes can be realized only in the samples of small sizes when the size of a sample~$L$ is less than a magnetic scale~$q $~\cite{K3}. In this case, the $q$ value  is  the maximum scale in the model.

IQHE is explained in the framework of the Hofstadter model~\cite{K1,K2,K3,a3,K4,B1,B2}, while the same cannot be said about FQHE. Unfortunately, the Hofstadter model is incapable of explaining FQHE, because it does not take into account the interaction between quantum particles.
FQHE is not sensitive to spin degrees of freedom, so the repulsion between fermions should be taken into account first.
A theory that could explain all the diversity of the FQHE is still lacking, and the nature of the FQHE remains an open question in condensed matter physics. Let us pay tribute to the ideas~\cite{A1,A2}, from which it becomes clear that the effect itself is not trivial.

The purpose of this work is not to explain the numerous experimental data on the measurement fractional Hall conductance, but to understand  the nature of the FQHE. The material of the paper is presented in the following format: original part as an example, well-known results as a counterexample.

\section{Model Hamiltonian and method}

We study FQHE in the framework of the Hofstadter model defined for interacting electrons on a square lattice with the Hamiltonian ${\cal H}={\cal H}_0  +{\cal H}_{\rm {int}}$
\begin{eqnarray}
{\cal H}_0&=& -\sum_{\sigma=\uparrow,\downarrow}\sum_{n,m} \left[a^\dagger_{n,m;\sigma} a_{n+1,m;\sigma} + \re^{2\ri \piup n\phi} a^\dagger_{n,m;\sigma}  a_{n,m+1;\sigma} + H.c. \right] \nonumber \\
 && -\mu\sum_{\sigma=\uparrow,\downarrow}\sum_{j} n_{j;\sigma}-H \sum_{j} \left(n_{j;\uparrow}-n_{j;\downarrow}\right),
 \label{eq-H0} \\
{\cal H}_{\rm {int}} &=&U \sum_{j} n_{j;\uparrow} n_{j;\downarrow},
 \label{eq-H1}
\end{eqnarray}
where $a^\dagger_{n,m;\sigma} $ and $a_{n,m;\sigma}$ are the fermion operators located at a site $j=\{n,m\}$ with spin $\sigma =\uparrow,\downarrow$, $n_{j;\sigma}=a^\dagger_{j;\sigma}a_{j;\sigma}$ denotes the density operator,
$\mu$ is a chemical potential. The Hamiltonian ${\cal H}_0$ describes the hoppings of fermions between the nearest-neighbor lattice sites. A magnetic flux through the unit cell $\phi = {H }/{ \Phi_0}$ is determined in the quantum flux unit ${\Phi_0=h/e}$. Here, $H$ is a magnetic field and  a lattice constant is equal to unit. ${\cal H}_{\rm {int}}$ term is determined by the on-site Hubbard  interaction $U$.

The interaction term~\eqref{eq-H1} can be conveniently redefined in the momentum representation
${\cal H}_{\rm {int}} = V U  \sum_{\textbf{K}}n_{\textbf{K};\uparrow} n_{-\textbf{K};\downarrow}$,
where $n_{\textbf{K};\sigma}=\frac{1}{V}\sum_j \exp(\ri\textbf{K j})n_{j;\sigma}$, the volume is equal to $V=L \times L$. Using the mean field approach, we rewrite this term as follows ${\cal H}_{\rm {int}} =V(\lambda_{\textbf{K};\uparrow}n_{\textbf{-K};\downarrow}+\lambda_{\textbf{-K};\downarrow}n_{\textbf{K};\uparrow})$ with an effective field $\lambda_{\textbf{K};\sigma}=U \langle n_{\textbf{K};\sigma} \rangle$, which is determined by a fixed value of the wave vector $\textbf{K}$.  In this case, the value of  $\textbf{K}$ is a free parameter of the mean-field approximation that minimizes the energy of the electron liquid, in contrast to $q$ the value of which is determined by an external magnetic field.
In the experiments, the magnetic fields  correspond to the semi-classical limit with a magnetic scale $q \sim 10^3$--$10^4$, which corresponds to small values $K\sim 10^{-3}$--$10^{-4}$. The density of fermions for the states near the low energy edge of  the spectrum is small $\sim {1}/{q}$.
In the small $\textbf{K}$-limit, the expression for $\lambda_{\textbf{K};\sigma}$ is simplified  $\lambda_{\textbf{K};\sigma}=\lambda_\sigma +0(K^2)$, where $\lambda_\sigma =U\rho_\sigma $, $\rho_\sigma$ is the density of electrons with spin $\sigma$.
The Zeeman energy shifts the energies of electron bands with different spins, removes the spin degeneracy, and does not change the topological state of the electron liquid. This makes it possible to explicitly disregard the dependence of the electron energy on the spin and to consider the problem for spinless fermions. The model is reduced to a spinless fermion liquid with the interaction term  
$${\cal H}_{\rm {int}} =\frac{\lambda}{2}\sum_{j}\left[\exp(\ri\textbf{K j})+\exp(-\ri\textbf{K j})\right]n_j=\lambda \sum_{j}\cos(\textbf{K j})n_j,$$ 
with $\lambda =U\rho$, $n_j$ and  $\rho$ are the density operator of spinless fermions and their filling.

We study the 2D system in a hollow cylindrical geometry with open boundary conditions (a cylinder axis along the $x$-direction and the boundaries along the $y$-direction). The Hamiltonian ${\cal H}_0 $ describes the chains of spinless fermions oriented along the $y$-axis ($n$ is a coordinate of the chain in the $x$-direction) connected by single-particle tunneling with the tunneling constant equal to unit. The wave function of free fermions in the $y$-chains, which is determined by the wave vector $k_y$ are localized in the $x$-direction~\cite{1,3} (for each $k_y$ value). The amplitudes of the wave function with different values $k_y$ overlap in the $x$-direction, the eigenstates of the Hamiltonian ${\cal H}_0 $  are the Bloch form. All states with different $n$ are bounded via a magnetic flux. The on-site Hubbard interaction does not break the time reversal symmetry and chirality of the spectrum of the fermion liquid. Therefore, the effective Hamiltonian also should not break these symmetries for rational fluxes. These conditions are fulfilled in the case when $\textbf{K}=(K,0)$, where $K$ and $q$ form the states with rational periods. Making the ansatz for the wave function $\psi(n,m)=\exp(\ri k_y m)g_n$ (which determines the state with the energy $\epsilon$), we obtain the Harper equation for the model Hamiltonian~\eqref{eq-H0},~\eqref{eq-H1}
\begin{equation}
\epsilon g_n =-g_{n+1}-g_{n-1}-2\cos(k_y +2\piup n \phi)g_n +\lambda \cos(K n)g_n.
 \label{eq-Har}
\end{equation}
This equation is the key in studying FQHE.

The problem is reduced to the ($1+1$)D quantum system, where the states of fermions are determined by two phases: the first is  a magnetic phase $\phi$, the second is the phase $K$, which is connected with interaction. The  $K$ value corresponds to the minimum energy of the system. It minimizes the energy of electron liquid upon interaction~\eqref{eq-H1}. At $T=0$~K, the model is a three parameter model. We shall analyze the phase state of the interacting spinless fermions for arbitrary rational fluxes $\phi={p}/{q}$ ($p$ and $q$ are coprime integers), $U$ and $\rho$.
In the Hofstadter model of noninteracting fermions, the states of fermions with different $\phi$ are topologically  similar in the following sense: the Chern numbers, the Hall conductance are determined by the magnetic flux, filling or the number of the filled isolated Hofstadter bands (HBs) that correspond to this filling, while they do not depend on the structure of the bands~\cite{K1,K2,K3} (their widths, the values of the gaps between them).

The effect of the interaction on the behavior of fermion liquid is reduced to the appearance of an inhomogeneous $\lambda$-field, which is determined by the magnitude $\lambda$ and phase $K$. We shall use the following parametrization  $K=2 \piup {r}/{s}$, where $r$ and $s$ are relatively prime integers. Such trivial solutions $K=0$ ($s\to\infty$) and $K=2\piup$ ($r=s=1$) correspond to the maximum energy. According to~\eqref{eq-Har}, the energy~$\epsilon$ is shifted to the maximum value $+\lambda$. In the $K \to 0$ (or $s\to\infty$) limit, the solution for $K $ corresponds to irrational fluxes that are realized at $s>L$~\cite{K3}. We consider the steady state of the system for rational fluxes, namely for integer ${s}/{q}=\alpha$, when $q \leqslant s$, or integer $\alpha^{-1}$, when $q>s$. The minimum energy corresponds to nontrivial solution for $K$ at a given magnetic flux $\phi$.
The fine structure of the Hofstadter spectrum is realized at $\alpha>1$, when the interaction scale is maximum $s>q$.
In the case $\alpha<1$, the spectrum is renormalized, its  structure remains the same, i.e., only the Landau levels.

\section{Example of explanation of FQHE}
\subsection{Splitting of low energy Hofstadter bands,
a fine structure of the spectrum in the semi-classical limit}
\label{sec:2}

\begin{figure}
\centering{\leavevmode}
    \includegraphics[width=6.7cm]{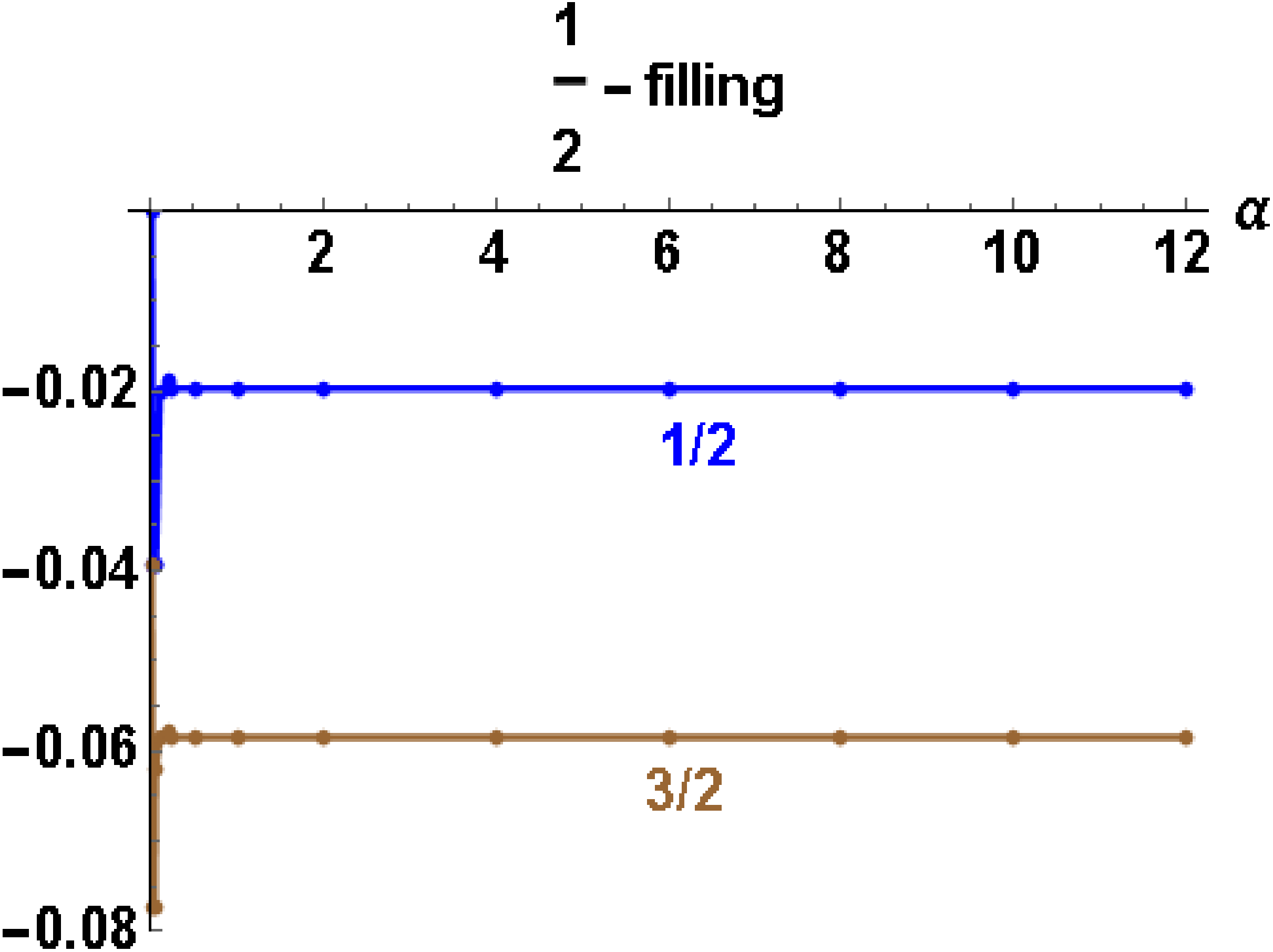} a)
    \hspace{0.2cm}
	\includegraphics[width=6.7cm]{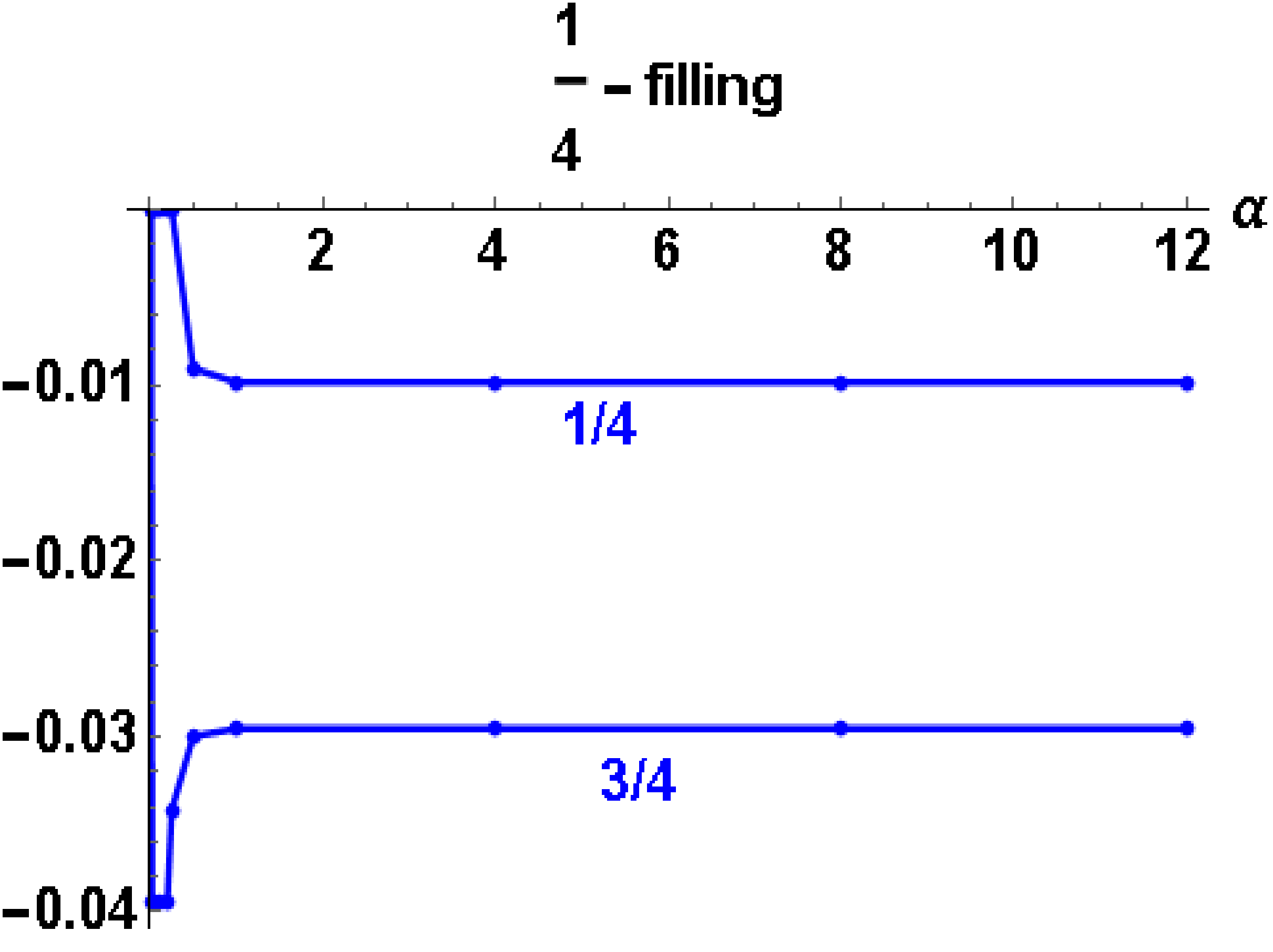} b)
	\hspace{0.2cm}
	\includegraphics[width=6.7cm]{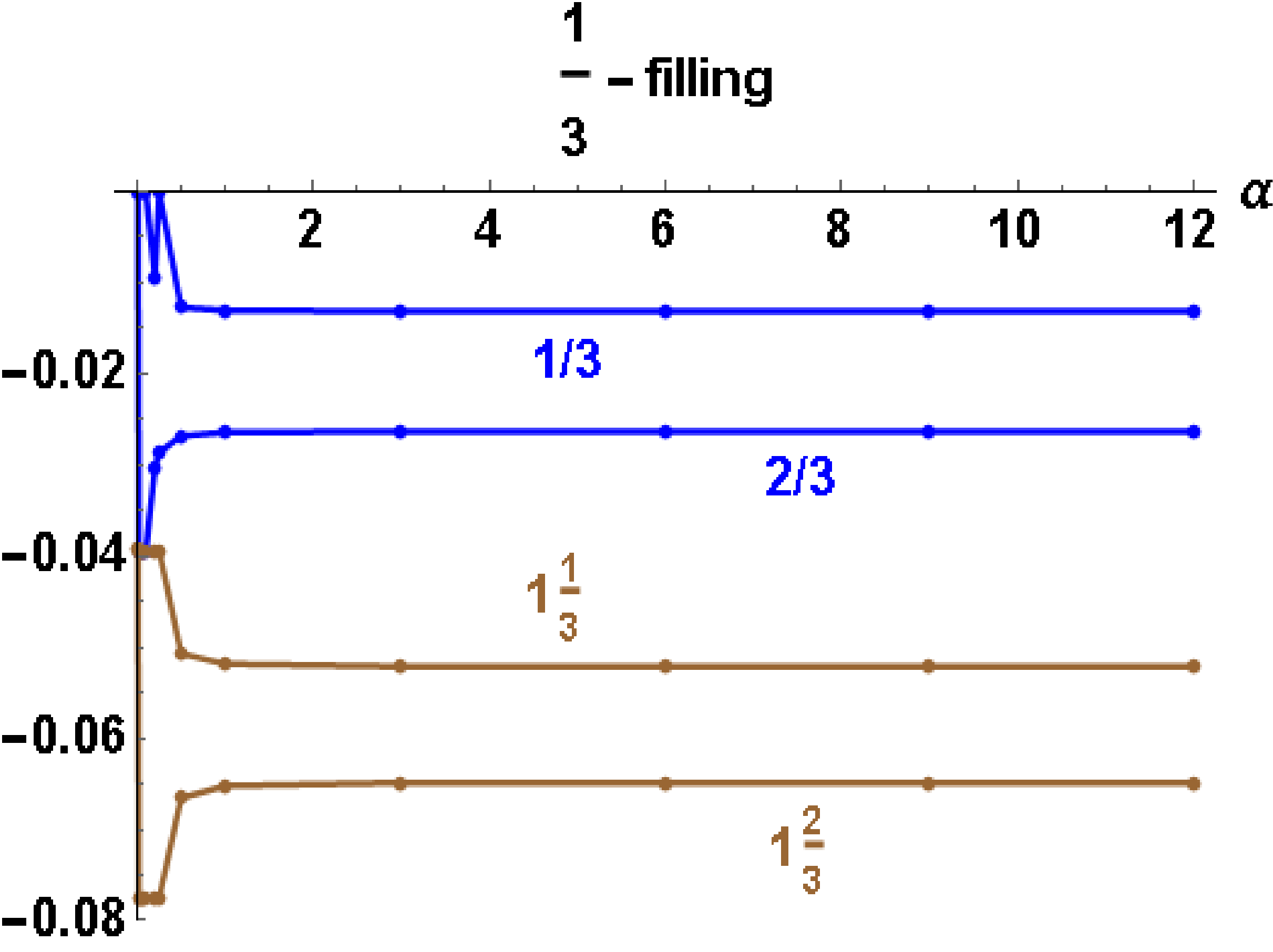} c)
	\hspace{0.2cm}
	\includegraphics[width=6.7cm]{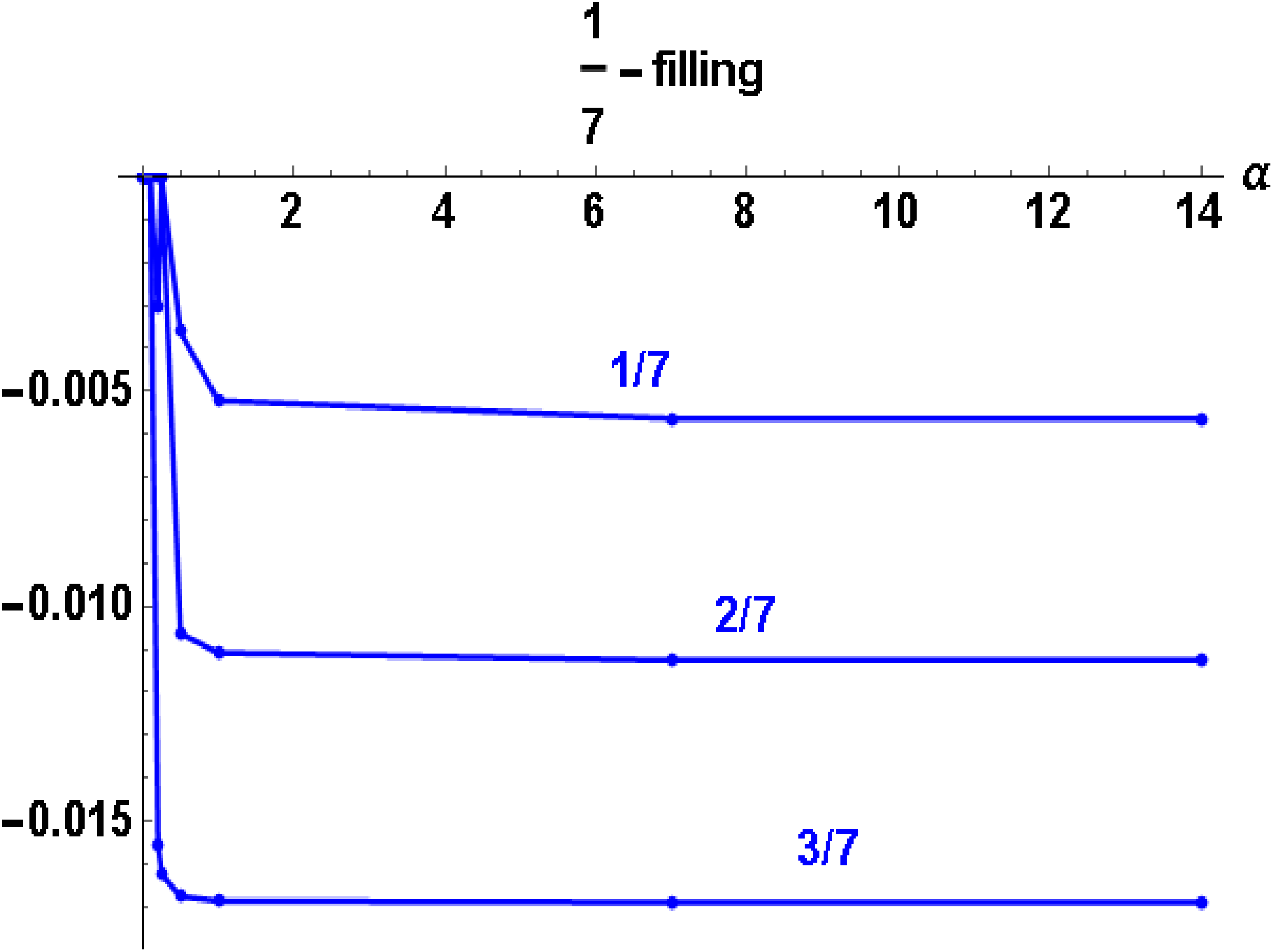} d)
	\hspace{0.2cm}
    \includegraphics[width=6.7cm]{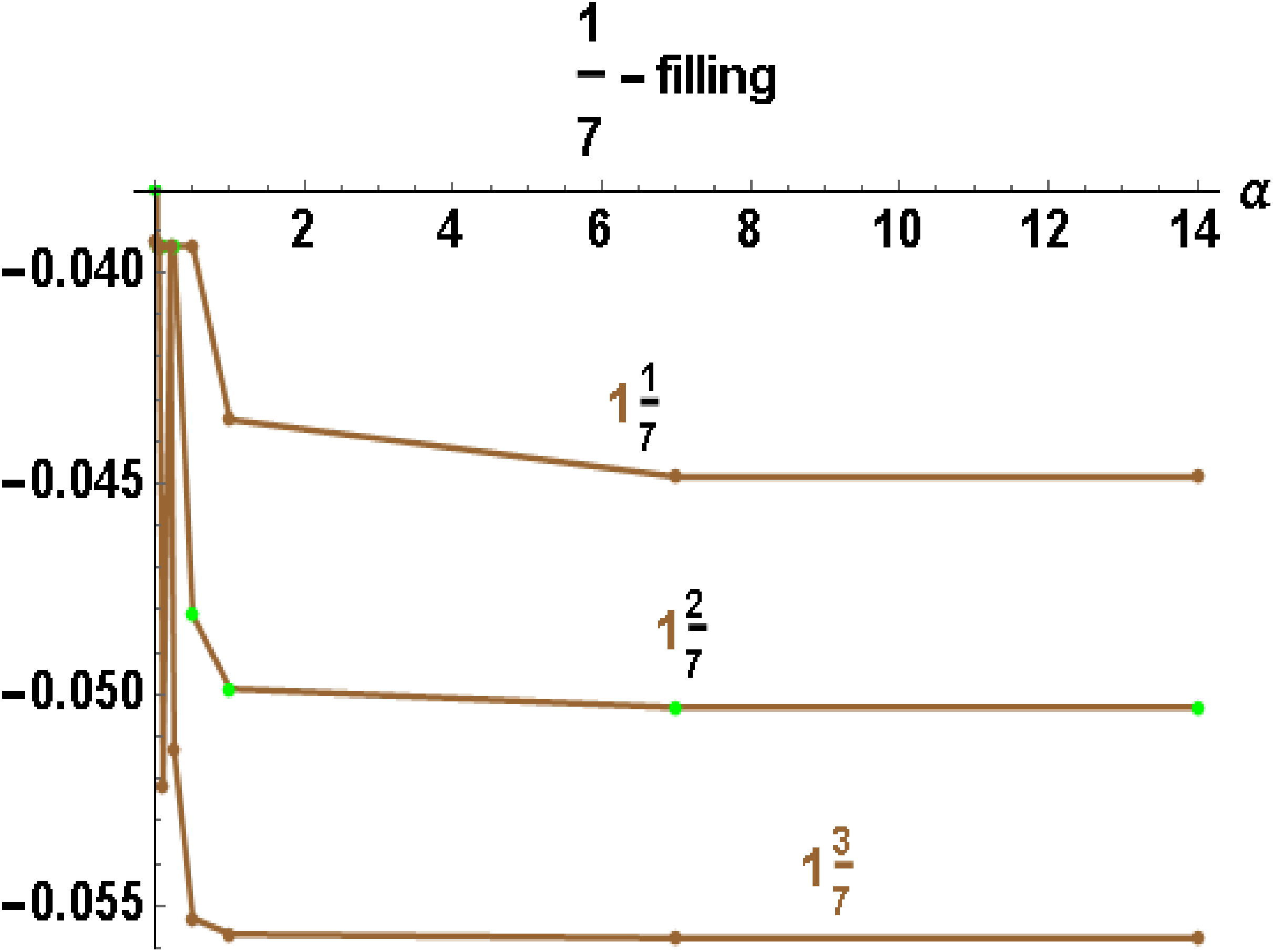} e)
	\hspace{0.2cm}
	\includegraphics[width=6.7cm]{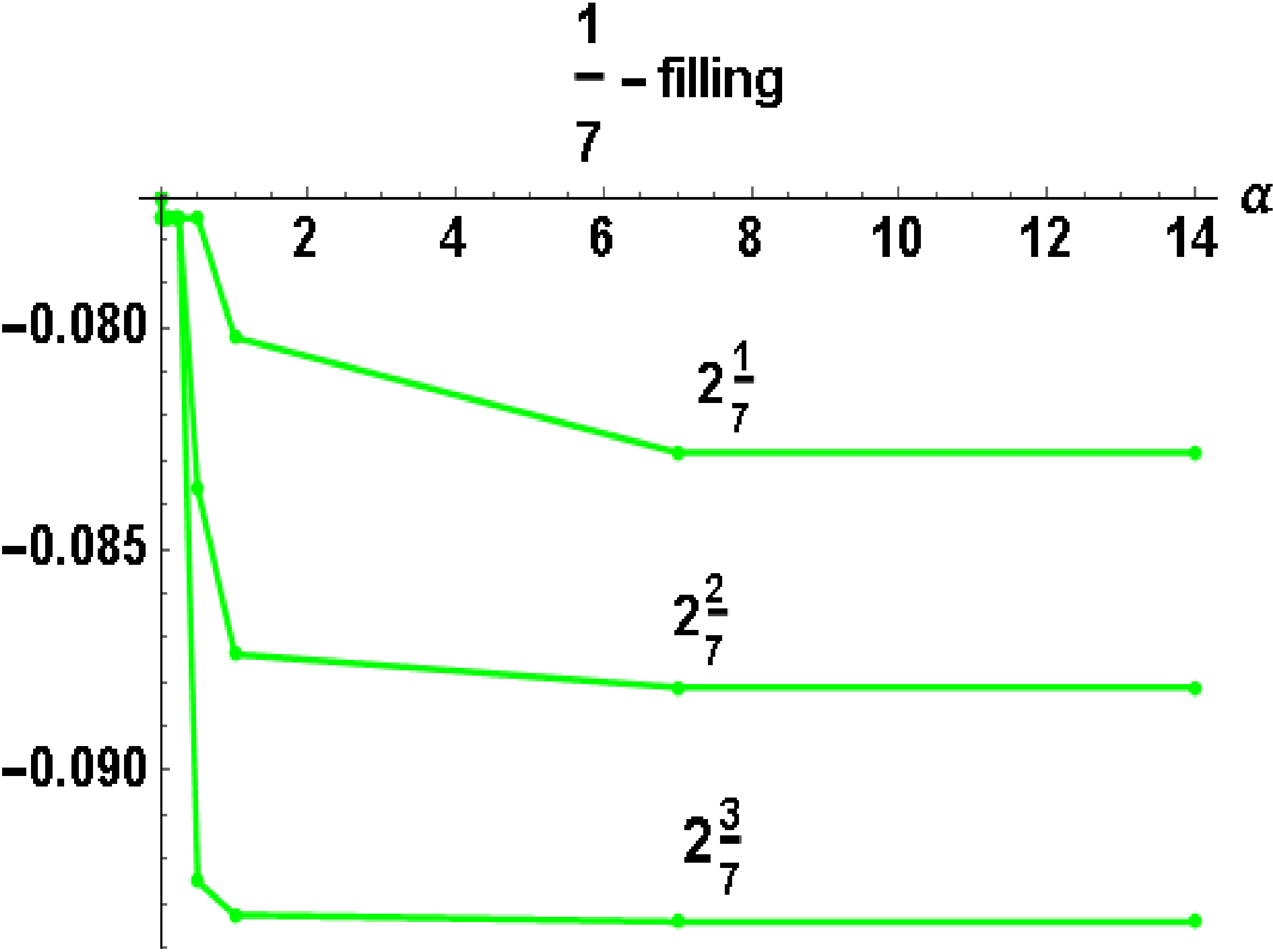} f)
\caption{(Colour online) The energy density as a function of $\alpha$ calculated at $q=10^2, U=1$, $\lambda =U(\rho_r+\nu/q)$ ($\rho_r$ is the fermion density corresponding to $r$-filled HB, $\nu$ is the  fractional filling of the $r+1$ HB) for:
a) $\frac{1}{2}$-filling for the first (blue line) and second (brown line) HB (the points $\alpha \geqslant 2 $ characterize the unstable fine structure of HB);
b) $\frac{1}{4}$-filling for the first HB  [$\frac{2}{4}$- or
$\frac{1}{2}$-filling is shown in a)], steady fractional state with $\alpha\geqslant 4$ and filling $\frac{1}{4}$ is also  realized in the second HB;
c)  $\frac{1}{3}$-filling for the first (blue lines) and second (brown lines) HB, steady fractional state for $\nu=\frac{1}{3}$  unsteady state  for $\nu=\frac{2}{3}$; $\frac{1}{7}$- filling for the first (blue lines) d), second (brown lines) e) and third (green lines) f) HB, the steady states are shown at $\nu =\frac{1}{7}, \frac{2}{7}, \frac{3}{7}$. }
       \label{fig:1}
\end{figure}

First of all we provide numerical analysis of the quasi-particle excitations near the edge of the spectrum considering rational fluxes $\phi$ and $K$   in the semi-classical limit with $p=1$ and $r=1$ for different $q\gg 1$, $s\gg 1$ and filling $\rho \ll 1$. Magnetic fields, at which measurements are carried out, correspond to large $q\sim 10^3$--$10^4$, so there is no point in considering the case of $q \sim 1$. As a reasonable compromise with numerical calculations (large, but not very large $q$), we consider the splitting (due to the interaction) of low energy fermion bands at $q=10^2$. For states near the edge spectrum, the value of $\lambda$ corresponds to a weak interaction limit, because the filling $\rho \sim {1}/{q}$ or $\rho \sim {1}/{s}$.

We show that the FQHE is determined by the fine structure of the spectrum, which is formed due to the on-site repulsion in an external magnetic field.  We consider the formation of a fine structure of low energy HBs, which correspond to filling less than $  {1}/ {q} $ for the first (the lowest) HB and when filling ${1} /{q} \leqslant\rho \leqslant {2} /{q} $ for the second band, where $ q = 10^2 $ is fixed for numerical calculations.
A rather obvious consequence follows from numerical calculations: in  a weak coupling at $\rho U <1$, that is valid in semi-classical limit for an arbitrary bare value of $U$, the fine structure of the spectrum does not depend on the value of $ \lambda $.  This allows us to consider the evolution of the fermion spectrum for a fixed value of $U=1$ or $\lambda=\rho$ and different $s$. We fix $ q =10^2$, $U=1$ and calculate the spectrum for various  $ \alpha = {s}/{q} $, which corresponds to rational fluxes, when $\alpha$ or $\alpha^{-1}$ is an integer.

It is really nice that the spectrum has a fairly simple topologically stable structure. The number of HBs in the spectrum is equal to $q$,  at $\alpha>1$,  $\alpha$ subbands form a fine structure of each HB.  The values of the gaps between low energy HBs $\Delta_{j,j+1}(\alpha)$ ($j$ numerates the band) depend on $q$ and $\lambda$ and insignificantly depend on $\alpha$. At $q=10^2$ and $U=1$, $\Delta_{1,2}(\alpha)\simeq 0.1038$, $\Delta_{2,3}(\alpha)\simeq0.082$ for $2\leqslant\alpha\leqslant 7$, and $\Delta_{1,2}=0.1237$, $\Delta_{2,3}=0.1217$ at $U=0$ in the Hofstadter model, for comparison (details of the calculation are presented in Appendix~\ref{A}) $\alpha$-narrow subbands form a fine structure of the $j$-HB, the bandwidth of $i$-subband in $j$-HB is denoted as $\epsilon_{j,i}(\alpha)$.
According to numerical calculations provided in Appendix~\ref{A} $\epsilon_{j,i}(\alpha)<0.02$ for $j=1,2$ and $1\leqslant\alpha\leqslant7$, their values increase with an increase of the HB number and decrease with an increase of $\alpha$.

Extremely small quasigaps $\delta\varepsilon_{j;i,i+1}(\alpha)$ separate subbands $i$ and $i+1$ in the fine structure of the $j$-HB. Their values are calculated for two HBs  $\sim 10^{-10}$--$10^{-13}$ (the calculated values of $\delta\varepsilon_{j;i,i+1}(\alpha)$ are given in Appendix~\ref{A}). A fine structure of the spectrum forms from the Dirac subbands.

A structure of the spectrum which includes two low energy HB has the following form for $\alpha=3$ as an example:
\begin{eqnarray}
	&&\epsilon_{1,1}(3)=0.0017 \Longrightarrow \delta\varepsilon_{1;1,2}(3)\sim 5\cdot 10^{-12}\Longrightarrow\epsilon_{1,2}(3)=0.0067 \Longrightarrow\delta\varepsilon_{1;2,3}(3)\sim 9\cdot 10^{-11} \nonumber\\
	&&\Longrightarrow
	\epsilon_{1,3}(3)=0.0050 \Longrightarrow \Delta_{1,2}(3)= 0.1038 \Longrightarrow \epsilon_{2,1}(3)= 0.0066\Longrightarrow \delta\varepsilon_{2;1,2}(3)\sim 9\cdot 10^{-13} \nonumber\\
	&&\Longrightarrow \epsilon_{2,2}(3)= 0.0166\Longrightarrow \delta\varepsilon_{1;2,3}(3)\sim 9\cdot 10^{-11}\Longrightarrow \epsilon_{2,3}(3)= 0.0100\Longrightarrow\Delta_{2,3}(3)=0.0820.\nonumber
\end{eqnarray}
Positive values  of $\delta\varepsilon_{j;i,i+1}(\alpha)$  correspond to the quasigaps or zero density states at fraction fillings, since their values are extremely small $\sim 10^{-10}$--$10^{-13}$. In semiclassical limit, the fine structure of the low energy HB does not change  at different $q$.
According to numerical calculations,  at a given $\alpha$ each HB is split by  quasigaps into $\alpha$ subbands, the fine structure of each HB is formed.
Thus, the spectrum is determined by two types of gaps, the gaps $\Delta_{j,j+1}$ which determine the insulator states of the system with an entire filling $\rho_j ={j}/{q}$ (where $j$ is integer) and $\delta\epsilon_{j;i,i+1}$ with a fractional filling in each HB $\nu={i}/{\alpha}$ (here,  $i=1,\dots,\alpha-1$). Moreover, $\Delta_{j,j+1}\gg \delta\epsilon_{j,i;j,i+1}$. Thus, the structure of the spectrum is preserved in a fairly wide range of values $\lambda$, as noted above. Most likely, the quasigaps in the spectrum determine the points of tangency of the subbands, and therefore they are defined as quasigaps and the spectrum of HB includes only Dirac subbands.

\subsection{Fractional filled steady state}
\label{sec:1}
In this subsection we consider a stability of the fine structure of the Hofstadter spectrum. Let us fix the chemical potential which corresponds to the fractional filling of each HB (for $\alpha =2$, $\nu=\frac{1}{2}$) and numerically calculate the energy of electron liquid for different rational fluxes, which corresponds to integer~$\alpha$ and $\alpha^{-1}$. A steady state corresponds to the minimum energy for a given filling. The energy density as a function of~$\alpha$ is shown in figure~\ref{fig:1}~a). A steady state is realized at $\alpha=\frac{1}{20}$ for the first HB and $\alpha=\frac{1}{25}$ for the second. As a result, the fine structure of the Hofstadter spectrum is unstable at $\alpha=2$ or $\frac{1}{2}$-fractional filling.
It follows from numerical analysis of stability of fine structures at different $\alpha$ that the fine structure of HB is stable when HB is  filled $\nu< \frac{1}{2}$. The point $\nu =\frac{1}{2}$ is similar to the point of the phase transition. Therefore, in this point  the behavior of the electron liquid is rather critical. In figure~\ref{fig:1} we also presented the calculations of energy density for $\nu=\frac{1}{4},\frac{3}{4}$ for the first HB b),
$\nu=\frac{1}{3},\frac{2}{3}$  for the first and second HB c), $\nu=\frac{1}{7},\frac{2}{7}, \frac{3}{7}$ for the first d)  second e) and third f) HB.
For steady fractional Hall states, the minimum energy is reached at $\alpha_c=3$ for $\nu=\frac{1}{3}$, $\alpha_c=4$  for $\nu=\frac{1}{4}$, $\alpha_c=7$ for $\nu=\frac{1}{7},\frac{2}{7}, \frac{3}{7}$. For rational fluxes, the energy density does not depend on the value of $\alpha$ at $\alpha>\alpha_c$. The Hubbard interaction shifts  HB (decreasing the energy compared to the homogeneous state) and increases the bandwidths of subbands (increasing the energy). The summarized energy  occurs as a result of the competition of these terms, that are determined by $U$.

\subsection{The Dirac spectrum.\\
Edge modes, fractional Hall conductance}
\label{sec:2}
It is convenient to consider the behavior of the edge modes when calculating quasiparticle excitations for the stripe geometry with open boundary conditions: the boundaries, parallel to the $y$-axis, are the edges of the hollow cylinder.
Let us analyse the behavior of the fermion spectrum  in the case of a strong anisotropic hopping integral in equation~\eqref{eq-Har}
\begin{equation}
\epsilon g_n =-t g_{n+1}-t g_{n-1}-2\cos(k_y +2\piup n \phi)g_n +\lambda \cos(K n)g_n,
 \label{eq-Har2}
\end{equation}
where the hopping integral for fermions between chains $t\ll1$.

The behavior of the topological properties of the system is universal in the sense that they do not depend on the parameters of the Hamiltonian over a wide range of their variation. This makes it possible to analyze the spectrum of the quasi-particle excitations in a weak limit with respect to $t$. The Bloch fermion states in the chains are described by the excitation energies $\kappa (k_y,n)=-2\cos (k_y+2\piup \frac{n}{q})+\rho U\cos(2\piup/s)$ at $t=0$. At $t \neq 0$, the fermions tunnel between the chains. In the case when the energies of fermions in the chains $n_1$ and $n_2$ coincide, the interaction between fermions is maximum and is determined by a distance between the chains  $\sim t^{|n_1-n_2|}$ in the weak $t$-limit. For $U=0$, the fermion spectrum is gapped in these resonance points, while in the semi-classical limit it is described by the Landau levels. The tunneling of fermions in the gaps is determined as the tunneling between Majorana fermions with different chirality $\chi_n$ and $\nu_n$, located at the different chains~\cite{K1,K2}. At the same time, the tunneling between the Majorana fermions located at different edges is forbidden, and these chiral modes are free and localized at different edges. The number of these chiral modes defines the Hall conductance in  IQHE~\cite{K1,K2}.

The Hubbard interaction breaks a condition of the resonance for the given energies between all chains. The fermion energies in the chains are the same for only three chains. As a result, the gaps into HBs are not formed. Chiral modes are localized at the edges; these modes are defined by the conditions $\kappa (k_y,n_1)=\kappa(k_y,n_2)$, where $n_1$ and $n_2$ are the chains on different edges and $n_2=L-n_1+1$: for example, $n_1=1$ and $n_2=L$; $n_1=2$ and $n_2=L-1$. For $U>0$, the energy corresponding to the conditions $\kappa (k_y,1)=\kappa (k_y,2)=\kappa (k_y,L)$ is maximum, so the energies of the modes localized at the edges $1$ and $L$, split off from the upper edge of HB. The conditions $\kappa (k_y,n_1)=\kappa (k_y,n_2)$ for $n_1>1$ are realized for the corresponding energies inside HB. In other words, the edge mode moves inside the sample with a decreasing energy inside HB.
The energies at which the  chiral Majorana fermions are formed and localized at the edges with $n_1>1$ and $n_2<L$ correspond to the states inside HB.
The above is illustrated by numerical calculations of the fermionic spectrum.

\begin{figure}[h]
	\centering{\leavevmode}
	\includegraphics[width=6.9cm]{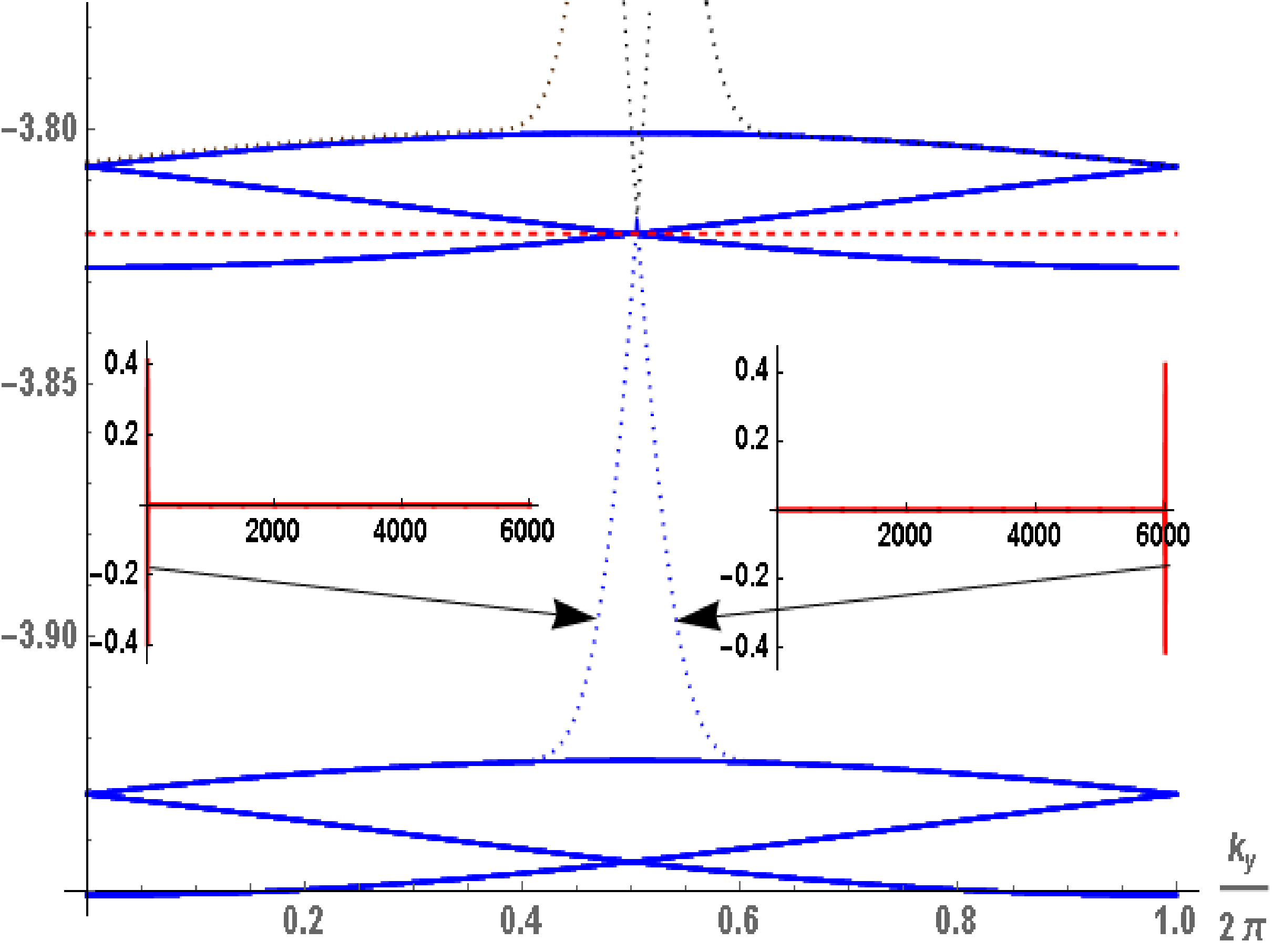} a)
	\hspace{0.2cm}
	\includegraphics[width=6.9cm]{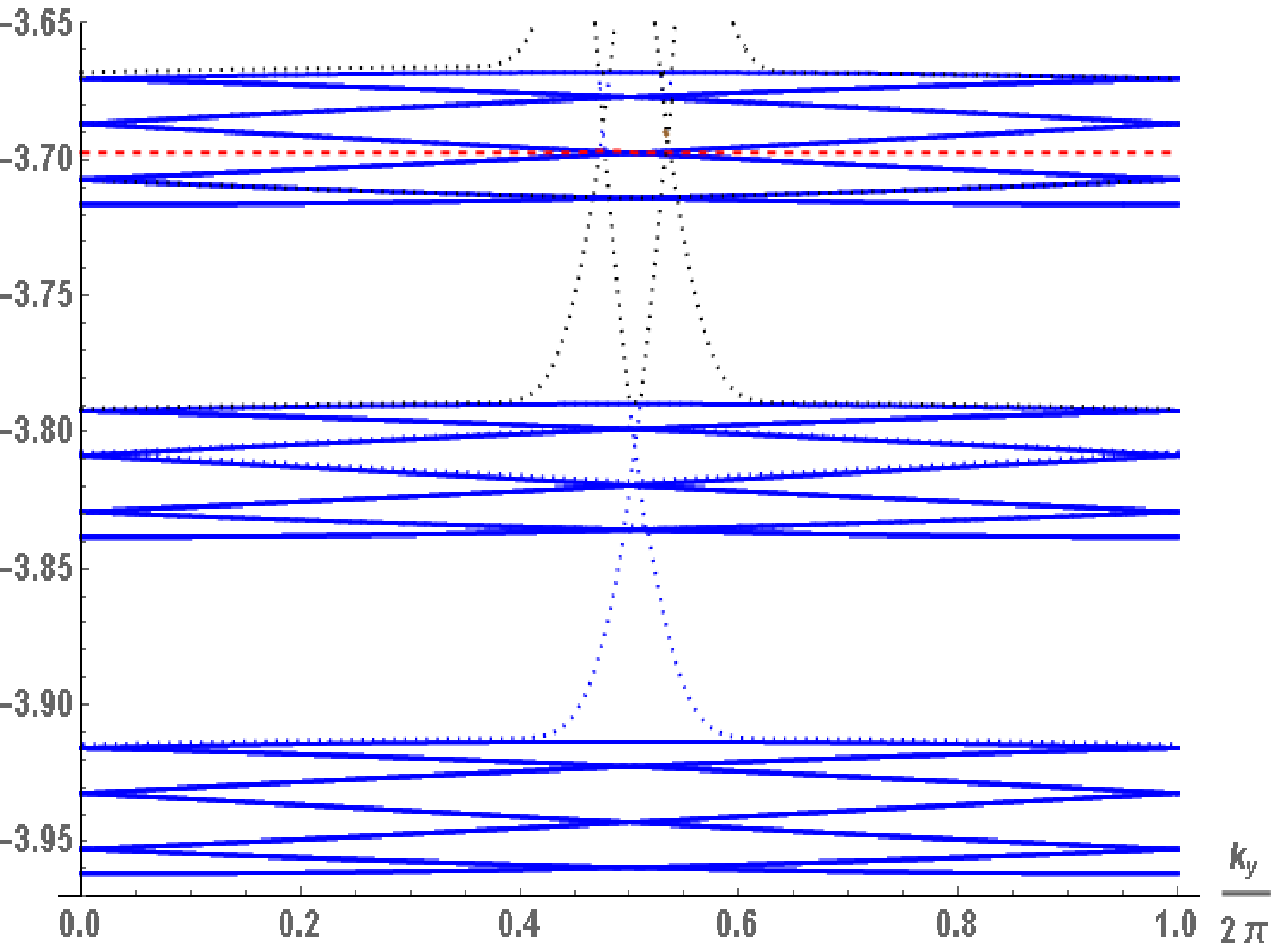} b)
	\caption{(Colour online)
		A fine structure of the two  a) and three b) lower energy HBs  (as illustration of the Dirac spectrum) calculated at $q = 100$, $U=1$, $\nu=\frac{1}{3}$ a) and $\nu=\frac{3}{7}$ b) for sample in the form of a  hollow cylinder with open boundary conditions along the $y$-direction, $k_y$ is the wave vector, red dashed lines denote the Fermi energies.
		The dotted lines mark the dispersion of edge modes, the inserts illustrate them, where the amplitude of the wave function is calculated as a function of the $x$-coordinate at ${k_y}/{2\piup}=0.48,0.52$ a)  (1 and $6 \cdot 10^3$ are the boundaries).
	}
	\label{fig:2}
\end{figure}

Let us focus on the calculation of two low energy HB at $\alpha=3$ and three HB at $\alpha=7$. We fix the Fermi energies which correspond to fractional filling $\nu=\frac{1}{3}$  in the second HB and $ \nu =\frac{3}{7}$ in the third HB. In the case $\alpha=3$, each HB splits into three subbands forming its fine structure. The quasigaps between their subbands are extremely small [see in figure~\ref{fig:2}~a)].  HB form the edge modes in the forbidden region of the spectrum between them. These modes split from the upper and lower HB and are localized at the boundaries. The edge modes coexist with the fine structure of each HB, except the first, in which they are not formed. In contrast to IQHE, these modes also connect the nearest-neighbor subbands in the fine structure of each HB (except the first one). Extremely small quasigaps do not kill the topology of HB (the number of the edge modes is conserved at a filling of HB).
The Hall conductance is determined by the same edge modes with different fraction filling $1+\frac{1}{3}$ for $\alpha =3$ and $2+\frac{3}{7}$ for $\alpha =7$.
Note that in the semi-classical limit the Dirac spectrum of the fermion excitations is realized for arbitrary fractional filling.

\newpage

\section{Conclusions}
The Hofstadter model with short-range repulsion is considered within the mean-field approach, which allows one to study FQHE.
We described IQHE and FQHE using the same approach and it is shown that:
\begin{itemize}
	\item short-range repulsion forms a steady fine structure of the Hofstadter spectrum when the filling of HB is less than a half;
	\item at fractional  filling of HB a fine structure of HB is formed from the Dirac subbands;
	\item these quasigaps  do not destroy the HB topology, only the HB determines the number of the edge modes (the Chern number of the HB is conserved);
	\item chiral edge modes located at the boundaries connect the nearest-neighbor subbands and determine the Hall conductance with fractional filling;
	\item chiral edge modes are not formed in the first HB. Therefore, fractional Hall conductance is not realized for the filling of the lowest (first) HB.
\end{itemize} 

A fine structure of the Landau levels (HBs) splits into the Dirac fermion spectra. For the half-filled Landau level, the Dirac composite fermions
were proposed in~\cite{KP1} and were studied in~\cite{KP2} in the framework of low-energy effective field theory.
Numerical calculations were carried out in the semi-classical limit, which corresponds to the experimental conditions.
The results obtained cannot be explained within the framework of the model and within the approach to its solution proposed in~\cite{ad}.





\appendix
\section{Example}\label{A}

Numerical calculations of the low energy structure of the spectrum are presented in this section, results of calculations were obtained  at fixed $q=10^2$ and  $U=1$. In the semi-classical limit, the Hofstadter spectrum is reduced to the Landau levels, which are separated by the gaps $\Delta_{j,j+1}(\alpha)$, where $j$ numerates HB and $\alpha$ determines the splitting of the band. Below we provide a set of the calculated values of~$\Delta_{j,j+1}(\alpha)$:
\begin{eqnarray}
\Delta_{1,2}(1)= 0.1043, \qquad\Delta_{2,3}(1)=0.0842;\nonumber\\
\Delta_{1,2}(2)= 0.1039,\qquad \Delta_{2,3}(2)=0.0824;\nonumber\\
\Delta_{1,2}(3)= 0.1038,\qquad \Delta_{2,3}(3)=0.0820;\nonumber\\
\Delta_{1,2}(4)=0.1037, \qquad \Delta_{2,3}(4)=0.0819;\nonumber\\
\Delta_{1,2}(5)=0.1037,\qquad \Delta_{2,3}(5)=0.0818; \nonumber\\
\Delta_{1,2}(6)=0.1037,\qquad \Delta_{2,3}(6)=0.0818; \nonumber\\
\Delta_{1,2}(7)=0.1037, \qquad\Delta_{2,3}(7)=0.0818. \nonumber
\end{eqnarray}
At $\alpha \geqslant 2$ the gap between HB is practically independent of $\alpha$.

A fine structure of HB is determined by the value of $\alpha$, each $j$-HB includes $\alpha$ subbands with band width $\epsilon_{j,i}(\alpha)$, where $i$ numetares the subband in the HB $1\leqslant i\leqslant\alpha$. The calculation results for two low-energy HBs are presented below 
\begin{eqnarray}
	&&\epsilon_{1,1}(1)=0.0197, \quad \epsilon_{2,1}(1)=0.0381;\nonumber\\
	&&\epsilon_{1,1}(2)=0.0050, \quad\epsilon_{1,2}(2)= 0.0100, \quad \epsilon_{2,1}(2)=0.0148, \quad \epsilon_{2,2}(2)= 0.0198;\nonumber\\
	&&\epsilon_{1,1}(3)=0.0017, \quad \epsilon_{1,2}(3)=0.0067 , \quad \epsilon_{1,3}(3)=0.0050, \quad
	\epsilon_{2,1}(3)= 0.0066, \quad\epsilon_{2,2}(3)= 0.0166,\nonumber\\
	&& \epsilon_{2,3}(3)= 0.0100; \nonumber\\
	&&\epsilon_{1,1}(4)=0.0007, \quad\epsilon_{1,2}(4)= 0.0035, \quad\epsilon_{1,3}(4)= 0.0053, \quad\epsilon_{1,4}(4)=0.0029,\quad \epsilon_{2,1}(4)=0.0036, \nonumber\\
	&&\epsilon_{2,2}(4)=0.0106, \quad\epsilon_{2,3}(4)=0.0123, \quad\epsilon_{2,4}(4)= 0.0058; \nonumber\\
	&&\epsilon_{1,1}(5)= 0.0004,\quad \epsilon_{1,2}(5)= 0.0020, \quad \epsilon_{1,3}(5)=0.0037,\quad \epsilon_{1,4}(5)=0.0040,\quad \epsilon_{1,5}(5)=0.0019,\nonumber\\
	&& 
	\epsilon_{2,1}(5)=0.0023, \quad\epsilon_{2,2}(5) =0.0070,  \quad\epsilon_{2,3}(5)=0.0099, \quad\epsilon_{2,4}(5)=0.0090, \quad \epsilon_{2,5}(5)=0.0038;\nonumber\\
	&&\epsilon_{1,1}(6)=0.0002,\quad \epsilon_{1,2}(6)=0.0012, \quad\epsilon_{1,3}(6)=0.0025, \quad\epsilon_{1,4}(6)= 0.0033,\quad \epsilon_{1,5}(6)= 0.0030, \nonumber\\
	 &&\epsilon_{1,6}(6)=0.0013, \quad\epsilon_{2,1}(6)=0.0016, \quad\epsilon_{2,2}(6)=0.0049,\quad\epsilon_{2,3}(6)=0.0075, \quad\epsilon_{2,4}(6)= 0.0083,
	 \nonumber\\
	&& \epsilon_{2,5}(6)= 0.0067, \quad\epsilon_{2,6}(6)=0.0027; \nonumber\\
	&&\epsilon_{1,1}(7)= 0.0001, \quad\epsilon_{1,2}(7)= 0.0008,  \quad\epsilon_{1,3}(7)= 0.0017, \quad\epsilon_{1,4}(7)= 0.0025, \quad\epsilon_{1,5}(7)= 0.0029,\nonumber\\
	 &&	\epsilon_{1,6}(7)=0.0024, \quad\epsilon_{1,7}(7)= 0.0010, \quad\epsilon_{2,1}(7)= 0.0011, \quad\epsilon_{2,2}(7)= 0.0036, \quad\epsilon_{2,3}(7)= 0.0057,\nonumber\\&& \epsilon_{2,4}(7)= 0.0070,
	\quad
	  \epsilon_{2,5}(7)= 0.0069, \quad\epsilon_{2,6}(7)=0.0051,  \quad\epsilon_{2,7}(7)= 0.0020.\nonumber
\end{eqnarray}

As expected, the bandwidth in $j$-HB decreases with $\alpha$ and increases with $j$.

Narrow subbands with bandwidths $\epsilon(j,i)\ll\Delta(j,j+1)$ form the fine structure of each HB. Quasigaps between subbands $i$ and $i+1$ in fine structure of $j$-HB $\delta\varepsilon_{j;i,i+1}(\alpha)$ are extremally small, so they are the following values for two low energy HBs
\begin{eqnarray}
&&\delta\varepsilon_{1;1,2}(2)\sim 3\cdot 10^{-11}, \quad \delta\varepsilon_{2;1,2}(2)\sim 2\cdot 10^{-10};  \nonumber\\
&&\delta\varepsilon_{1;1,2}(3)\sim 5\cdot 10^{-12},\quad \delta\varepsilon_{1;2,3}(3)\sim 9\cdot 10^{-11},\quad
\delta\varepsilon_{2;1,2}(3)\sim 9\cdot 10^{-13},\quad \delta\varepsilon_{2;2,3}(3)\sim 1\cdot 10^{-10}; \nonumber\\
&&\delta\varepsilon_{1;1,2}(4)\sim 4\cdot 10^{-12},\quad \delta\varepsilon_{1;2,3}(4)\sim 3\cdot 10^{-11},\quad
\delta\varepsilon_{1;3,4}(4)\sim 6\cdot 10^{-11},\quad \delta\varepsilon_{2;1,2}(4)\sim 3\cdot 10^{-11}, \nonumber\\
&&\delta\varepsilon_{2;2,3}(4)\sim 9\cdot 10^{-13}, \quad \delta\varepsilon_{2;3,4}(4)\sim 2\cdot 10^{-11};\nonumber\\
&&\delta\varepsilon_{1;1,2}(5)\sim 1\cdot 10^{-11},\quad \delta\varepsilon_{1;2,3}(5)\sim 1 \cdot 10^{-11},\quad \delta\varepsilon_{1;3,4}(5)\sim 2\cdot 10^{-11},\quad
\delta\varepsilon_{1;4,5}(5)\sim 1\cdot 10^{-11},\nonumber\\ 
&&\delta\varepsilon_{2;1,2}(5)\sim 4\cdot 10^{-11},\quad \delta\varepsilon_{2;2,3}(5)\sim 1\cdot 10^{-10},\quad \delta\varepsilon_{2;3,4}(5)\sim 3\cdot 10^{-11},\quad \delta\varepsilon_{2;4,5}(5)\sim 2\cdot 10^{-11}; \nonumber\\
&&\delta\varepsilon_{1;1,2}(6)\sim 1\cdot 10^{-11},\quad \delta\varepsilon_{1;2,3}(6) \sim 4\cdot 10^{-12},\quad \delta\varepsilon_{1;3,4}(6) \sim 3\cdot 10^{-11},\quad
\delta\varepsilon_{1;4,5}(6) \sim 3\cdot 10^{-11},\nonumber\\ 
&& \delta\varepsilon_{1;5,6}(6) \sim 3\cdot 10^{-11},\quad
\delta\varepsilon_{2;1,2}(6)\sim 2\cdot 10^{-11},\quad \delta\varepsilon_{2;2,3}(6) \sim 9\cdot 10^{-11},\quad
\delta\varepsilon_{2;3,4}(6) \sim 1\cdot 10^{-10},\nonumber\\ &&\delta\varepsilon_{2;4,5}(6) \sim 3\cdot 10^{-11},\nonumber\\
&&\delta\varepsilon_{2;5,6}(6) \sim 5\cdot 10^{-11};\nonumber\\
&&\delta\varepsilon_{1;1,2}(7)\sim 3\cdot 10^{-12},\quad \delta\varepsilon_{1;2,3}(7)\sim 3\cdot 10^{-11},\quad
\delta\varepsilon_{1;3,4}(7)\sim 8\cdot 10^{-12},\quad
\delta\varepsilon_{1;4,5}(7)\sim 4\cdot 10^{-11}, \nonumber\\ 
&&\delta\varepsilon_{1;5,6}(7)\sim 2\cdot 10^{-11},\quad \delta\varepsilon_{1;6,7}(7)\sim  1\cdot 10^{-11},\quad
\delta\varepsilon_{2;1,2}(7)\sim 1\cdot 10^{-11},\quad \delta\varepsilon_{2;2,3}(7)\sim 2\cdot 10^{-11},\nonumber\\
&&\delta\varepsilon_{2;3,4}(7)\sim 9\cdot 10^{-11}, \quad \delta\varepsilon_{2;4,5}(7)\sim 5\cdot 10^{-11},\quad
\delta\varepsilon_{2;5,6}(7)\sim 5\cdot 10^{-11},\quad \delta\varepsilon_{2;6,7}(7)\sim  3\cdot 10^{-11}.\nonumber
\end{eqnarray}
Positive values  of $\delta\varepsilon_{j;i,i+1}(\alpha)$  correspond to quasigaps or zero density states at fraction fillings, since its values are extremely small $\sim 10^{-10}$--$10^{-13}$. It follows from numerical calculations that, in the semi-classical limit, the fine structure of a low-energy HB does not change for different $ q $.

\section{Counterexample}
\label{B}
As a counterexample,  we analyze the results of the paper~\cite{ad}. The authors  considered a similar model, namely the Hofstadter model with  a periodic potential 
$U_1\cos  (2 \piup x/a)+U_2\cos  (2 \piup x/b)$, 
for both small and large $U_1/\omega$, $U_2/\omega$, where   
$\omega$ is the cyclotron frequency.  The problem is reduced to a solution of equation~\eqref{eq-Har} for rational magnetic fluxes determined as $p/q$. The authors believe that the two-dimensional periodic potential forms the gaps into HB and obtain the Diophantine equation for Chern numbers at fillings that correspond to these gaps. I quote:  ``If the Fermi surface is located in the $r$-th gap of the $N$-th Landau level the total Hall conductance is equal to $\sigma_H=(e^2/h)(t_r+N-1)$ [equation~(10)], with $t_r $ the solution of'' $r=s_r q+t_r p$ ($|s|\leqslant p/2$) [equation~(9) in~\cite{ad}] . The Diophantine equation~(9) and formula~(10) are the main result of~\cite{ad}. The problem is reduced to the traditional Hofstadter model with the same Diophantine equation~(11). Therefore, the Hofstadter model with the two-dimensional periodic potential did not describe the fractional Hall states.

\begin{figure}[h]
	\centering{\leavevmode}
	\includegraphics[width=6.9cm]{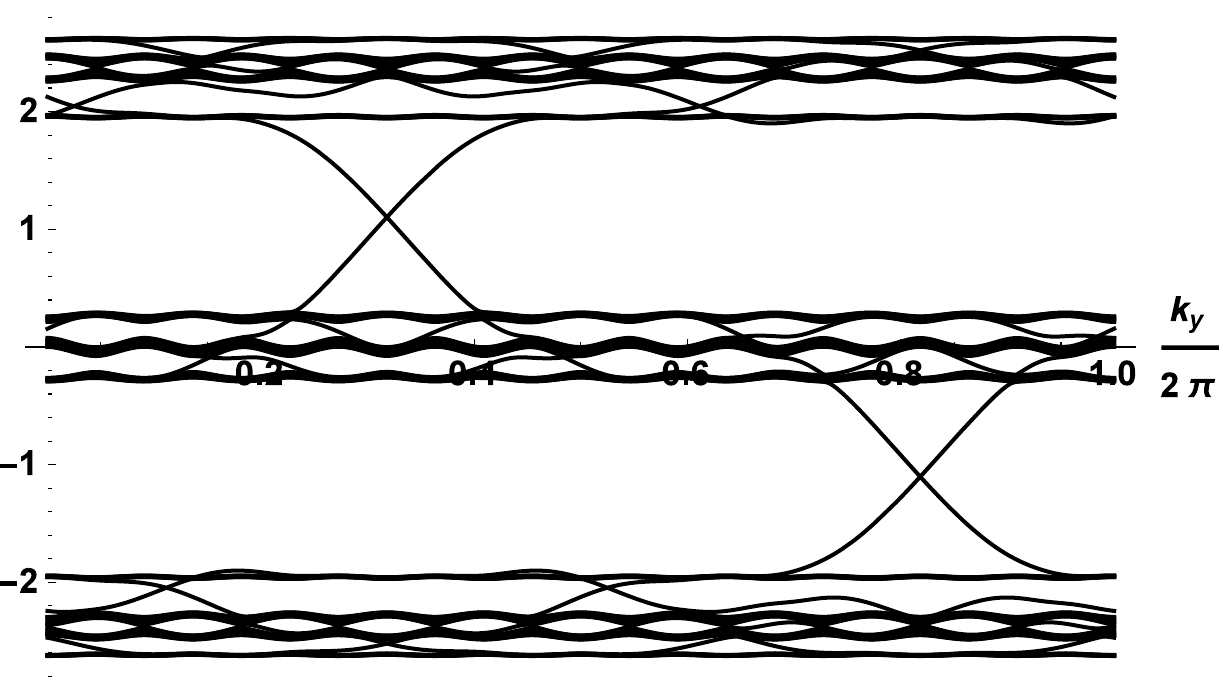} a)
	\hspace{0.2cm}
	\includegraphics[width=6.9cm]{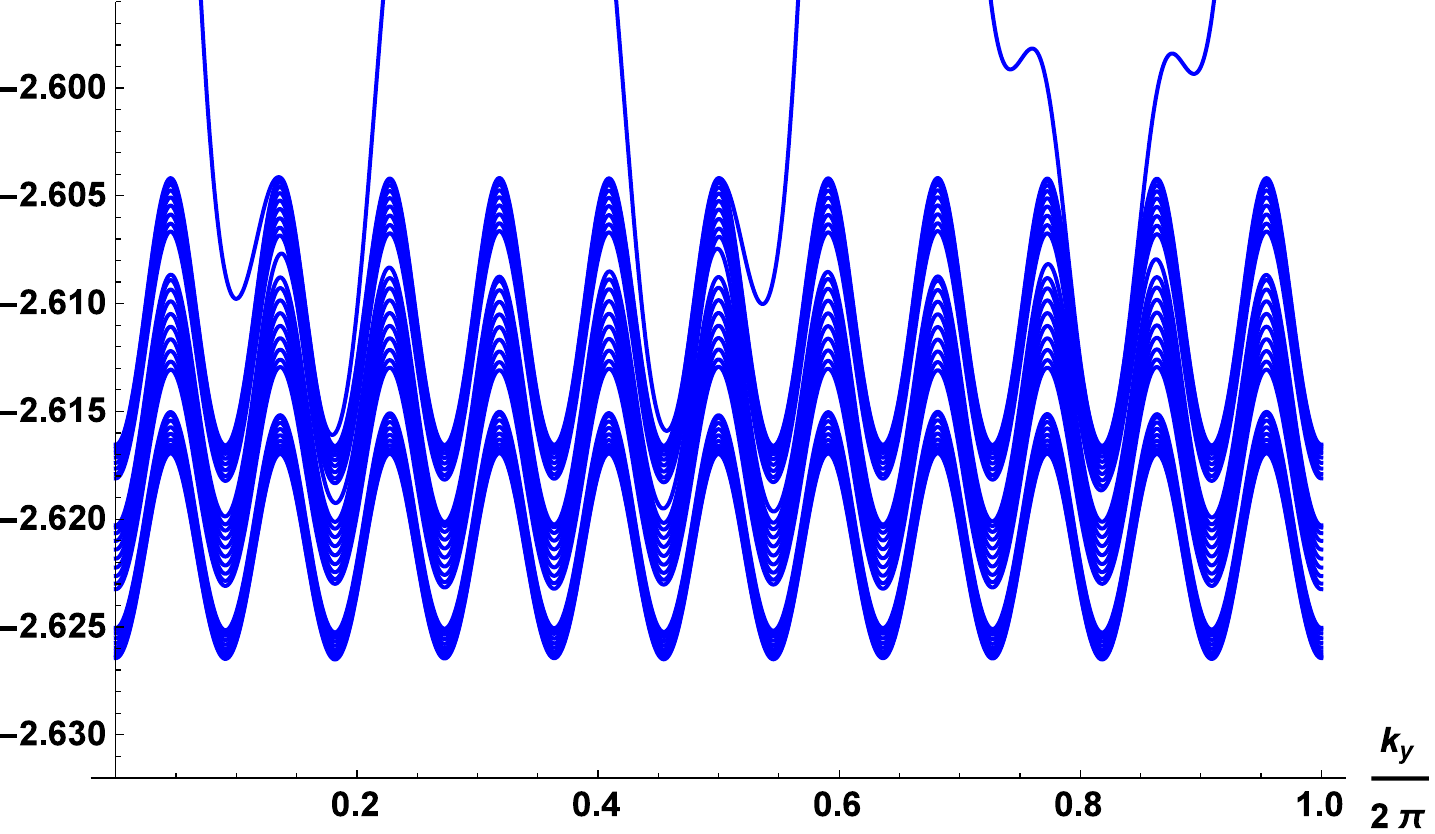} b)
	\caption{(Colour online) The Hofstadter spectrum calculated at $p=7$, $q=11$, $U_1=U_2=0$ a) and a fine structure of the lowest HB calculated at $p=7$, $q=11$,  $a=3 q$, $U_1=0.002$, $U_2=0$ b) for a sample in the form of a hollow cylinder with open boundary conditions along the $y$-direction, $k_y$ is the wave vector. }
	\label{fig:3}
\end{figure}

As an illustration, the authors considered  the flux to be equal to $\frac{7}{11}$ with the first 11 values of $t_r$: $-3$, 5, 2, $-1$, $-4$, 4, 1, $-2$, 6, 3, 0 and so that the Hall current is proportional to $-3$ or 8 in each subband. Unfortunately, this set of $t_r$  values does not satisfy the spectrum symmetry, such as $-3+5+2-1-4+4+1-2+6+3+0=11=q$. This spectrum is shown in figure~\ref{fig:3}~a) for $U_1=U_2=0$, the structure of the first HB is shown in figure~\ref{fig:3}~b) for $U_1=0.002$, $U_2=0$  and $a=3q=33$. Numerical analysis shows that the gaps that form the fine structure of each HB (see in figure~\ref{fig:3}) are absent for a weak periodic potential and we cannot talk about the Hall conductance.

Numerical analysis (see the calculations obtained in the semi-classical limit) shows the absence of the gaps (only extreme small quasi-gaps or peculiarities of the density of states at partial filling), when the Hubbard interaction is taken into account. Note that there are no calculations of the gap values~\cite{ad}, only an assumption.

According to equations~(9), (10)~\cite{ad}, the Chern number is different for different HB fillings, and when the HB filling changes, the topological phase transitions occur between topological states with different topological indices. This result follows from  equations~(9), (10)~\cite{ad}. The periodic potential does not break the time reversal symmetry (like a magnetic field), it has a different nature and cannot induce topological phase transitions between  the states with different topological indices. Numerical calculations show that the Chern numbers of HB are not changed (in the sense that the number of chiral edge modes remains the same when the HB is filled). The obtained results really make sense, but not the topological states discussed in~\cite{ad}.


	\ukrainianpart

\title
{Ферміонний спектр Дірака дробових квантових холлівських станів
}
\author{І. М. Карнаухов}
\address{Інститут металофізики ім. Г. В. Курдюмова НАН України, бульвар Академіка Вернадського 36, 03142 Київ, Україна  
}
\makeukrtitle

\begin{abstract}
	Застосовуючи уніфікований підхід, ми досліджуємо цілочисельний квантовий ефект Холла і дробовий квантовий ефект Холла в моделі Гофштадтера з короткодіючою взаємодією між ферміонами. Ефективне поле, яке враховує взаємодію між ферміонами, визначається як амплітудою, так і фазою. Його амплітуда пропорційна силі взаємодії, фаза відповідає мінімальній енергії. Фактично задача зводиться до рівняння Гарпера з двома різними масштабами: перший --- магнітний масштаб з розміром комірки, що відповідає одиничному квантовому магнітному потоку, другий масштаб визначає неоднорідність ефективного поля, формує сталу тонку структуру поля спектра Гофштадтера і призводить до реалізації дробових квантових холлівських станів. У зразку скінченного розміру з відкритими граничними умовами тонка структура спект\-ра Гофштадтера складається з діраківських гілок ферміонних збуджень, а також включає тонку струк\-туру крайових хіральних мод.	
	Числа Черна топологічних зон Гофштадтера зберігаються під час формуван\-ня їхньої тонкої структури. Крайові моди формуються в зонах Гофштадтера, вони з'єднують найближчі сусідні підзони і визначають провідність для дробового заповнення.

	\keywords 
	цілочисельний квантовий ефект Холла, 
	дробовий квантовий ефект Холла,
	модель Гофштадтера
\end{abstract}

\end{document}